\documentclass[twocolumn]{jpsj2} 
\newcommand{\nc}{\newcommand}
\nc{\rnc}{\renewcommand}
\nc{\nn}{\nonumber}
\nc{\ch}{\cosh}
\nc{\sh}{\sinh}
\nc{\sech}{{\rm sech}}
\rnc{\Im}{{\rm{Im}\,}}
\rnc{\Re}{{\rm{Re}\,}}
\def\i{{\rm i}}
\def\e{{\rm e}}
\nc{\mfa}{{\mathfrak{a}}}
\nc{\mfab}{\overline{\mfa}}
\nc{\mfA}{{\mathfrak{A}}}
\nc{\mfAb}{\overline{\mfA}}
\nc{\mfb}{{\mathfrak{b}}}
\nc{\mfbb}{\overline{\mfb}}
\nc{\mfB}{{\mathfrak{B}}}
\nc{\mfBb}{\overline{\mfB}}
\nc{\mfsl}{{\mathfrak{sl}}}
\nc{\db}{\displaybreak[0]\\}
\nc{\bra}{\langle}
\nc{\ket}{\rangle}
\nc{\vf}{v_{\rm F}}
\nc{\Dth}{D_{\rm th}}
\nc{\Ds}{D_{\rm s}}
\nc{\tr}{{\rm Tr}}
\nc{\J}{\mathcal{J}}
\nc{\Jq}{\J_{\rm Q}}
\nc{\Je}{\J_{\rm E}}
\nc{\Js}{\J_{\rm s}}
\nc{\lamr}{\Lambda_{\rm R}}
\nc{\laml}{\Lambda_{\rm L}}
\nc{\ep}{\varepsilon}
\rnc{\H}{\mathcal{H}}
\nc{\M}{\mathcal{M}}
\rnc{\(}{\left(}
\rnc{\)}{\right)}
\rnc{\d}{{\rm d}}
\nc{\lam}{\lambda}
\nc{\Lam}{\Lambda}
\nc{\dprime}{\prime\prime}

\def\runauthor{Kazumitsu Sakai and Andreas Kl\"umper}

\title{Non-dissipative Thermal Transport and Magnetothermal 
Effect \\for the Spin-1/2 Heisenberg Chain}

\author{Kazumitsu \textsc{Sakai}$^{1}$\thanks{E-mail address:
sakai@stat.phys.titech.ac.jp} and Andreas \textsc{Kl\"umper}$^{2}$
\thanks{E-mail address: kluemper@physik.uni-wuppertal.de}}

\inst{$^{1}$Department of Physics, Tokyo Institute of Technology, 
            Tokyo 152-8551, Japan\\
      $^{2}$Bergische Universit\"at Wuppertal, Fachbereich Physik, 
         D-42097 Wuppertal,Germany}
\abst{Anomalous magnetothermal effects are discussed in the spin-1/2 
Heisenberg chain. The energy current is related to one of the non-trivial 
conserved quantities underlying integrability and therefore both 
the diagonal and off diagonal dynamical correlations of spin
and energy current diverge.
The energy-energy and spin-energy current correlations at finite 
temperatures are exactly calculated by a lattice path integral 
formulation. The low-temperature behavior of the thermomagnetic 
(magnetic Seebeck) coefficient is also discussed. Due to effects 
of strong correlations, we observe the magnetic Seebeck coefficient 
changes sign at certain interaction strengths and magnetic fields.}

\kword{transport properties, thermal conductivity, magnetothermal effect,
Seebeck coefficient, Heisenberg chain}

\begin{document}
\maketitle
%
%
In the last  two decades, 1D strongly correlated systems have 
attracted immense theoretical and experimental interest. 
One of the reasons stems from their unusual static and dynamical 
properties peculiar to 1D systems.
The spin-1/2 Heisenberg chain is one of the most fundamental  
solvable models for 1D magnetic insulators and has served as a 
testing ground for many approaches.

Recently, transport properties of low-dimensional strongly correlated 
quantum systems have been extensively studied from both theoretical
and experimental sides.\cite{ZP03}
Among them anomalously enhanced thermal conductivity 
\cite{Kudo,Sologubenko00,Hess01} and unconventional large spin 
diffusion constants\cite{Taki} have been reported in experiments 
on one- or quasi 1D materials with weak interchain interactions. 
These observations indicate the existence of non-dissipative 
transport properties in 1D quantum systems.
Theoretically, the existence of such anomalous properties has 
also been pointed out especially in quantum integrable 
systems.\cite{znp}
One of the criteria for anomalous transport is the existence of
a non-zero Drude weight.

In particular for the spin-1/2 XXZ chain, the Drude weight
for the spin transport has been obtained by Zotos for finite 
temperature at zero magnetic field.\cite{Zotos99} 
In the massless regime, the Drude weight is non-zero and hence the spin
transport is non-dissipative.  
In the massive regime without magnetic field a similar treatment yields zero
Drude weight at any temperature implying dissipative spin transport. This
feature, however, contradicts with numerical or field theoretical approaches
and is still debated.  \cite{NarMA98,AlGros1,FK03}

On the other hand,  the thermal Drude weight  $\Dth(T)$
for zero magnetic field has been recently obtained by the 
Bethe ansatz showing that the thermal transport is non-dissipative in 
both massless and massive regimes.
\cite{KS,SK} At low-temperature $\Dth(T)$ behaves as 
$\Dth(T)=\pi \vf T/3$ where $\vf$ is the velocity of excitations. 
This  universal behavior was also found in general systems in which 
the low-energy excitations are  described by a $c=1$ conformal field 
theory.\cite{MHCB,OCC03} In the massive regime, we find that 
$\Dth(T)\sim\exp(-\delta/T)/\sqrt{T}$ where $\delta$ is the one-spinon 
(respectively one-magnon) excitation gap for the antiferromagnetic 
(respectively ferromagnetic) regime.\cite{MHCB2}

In this article we will mainly discuss the thermal transport and 
magnetothermal effect 
in presence of finite magnetic fields. Because  the spin-reversal 
symmetry vanishes for finite fields,  the magnetothermal effect 
does appear.\cite{znp,LG03,MHCB3}
Consequently the thermomagnetic (magnetic Seebeck) coefficient is
finite. Moreover we observe the magnetic Seebeck coefficient 
changes sign for certain interaction strengths and magnetic fields, 
which can be interpreted as effects of strong correlations.

Let us consider the spin-1/2 Heisenberg XXZ chain with
magnetic fields $h$:
\begin{align}
   \H&=\sum_{k=1}^{L}h_{kk+1}-\frac{h}{2}\sum_{k=1}^{L}\sigma_k^z, \nn \\
    h_{kk+1}&=J\left\{\sigma_{k}^+\sigma_{k+1}^-
                     +\sigma_{k+1}^+\sigma_{k}^-
                     +\frac{\Delta}{2}(\sigma_k^z\sigma_{k+1}^z-1)
              \right\}.                                             \nn
\end{align}
Here we restrict ourselves to the critical antiferromagnetic regime 
$0\le\Delta\le 1$ and $J>0$. In this case, the anisotropy parameter 
$\Delta$ is conveniently parametrized by $\Delta=\cos\gamma$ 
($0\le\gamma\le\pi$).

The spin and energy current operator of the present system are written as
\begin{align}
    \Js=J\sum_{k=1}^{L}(\i\sigma_k^+\sigma_{k+1}^-+\text{h.c.}), 
\,
    \Je=\i J\sum_{k=1}^{L}[h_{k-1k},h_{kk+1}], \nn
\end{align}
respectively. 
The transport coefficients are determined by the Kubo formula
in terms of the correlation functions of the above defined 
current operators.
For instance, the thermal conductivity is determined by
\cite{Mahan}
\[
   \Re \kappa(\omega)=\frac{1}{T}\Re
                \left\{
                    L_{\rm QQ}-\frac{L_{\rm Qs}^2}{L_{\rm ss}}
                \right\}                      
    =\pi \Dth(T)\delta(\omega)+\kappa_{\rm reg}, 
\]
where 
\[
     L_{ij}=\int_0^{\infty}\d t \e^{-\i\omega t}
            \int_0^{\beta}\d\tau
            \bra \J_i(-t-\i\tau)\J_j \ket,
\,       \text{$\{i,j\}=\{\text{Q,s}\}$}. 
\]
Here the thermal current $\Jq$ should be $\Jq=\Je-h\Js$.

As already mentioned by Zotos in ref.~\citen{znp},
the energy current $\Je$ is a constant of motion
($[H,\Je]=0$)
and therefore the thermal Drude weight is finite
($\Dth(T)>0$) at all finite temperatures. 
Hence the thermal transport is non-dissipative.

For zero magnetic field $h=0$, the magnetothermal effect 
is always zero because  the system exhibits the spin-reversal 
symmetry. 
In contrast, for finite magnetic fields $h>0$ we see $\bra \Je\Js \ket>0$,
which leads to the 
diverging off diagonal dynamical correlations:
\[
\Re L_{\rm Qs}=\pi(\beta \bra\Je \Js \ket -h \Ds(T))\delta(\omega),
\,\,
L_{\rm sQ}=L_{\rm Qs}.
\]
In this case the thermal Drude weight
\begin{equation}
   D_{\rm th}(T)=\beta^2 \bra \Je ^2\ket-
                 \beta^3 \frac{\bra \Je \Js \ket^2}{D_{\rm s}(T)},
   \label{thermalDrude}
\end{equation}
and the magnetic Seebeck coefficient
\begin{equation}
   S(T)=\Re \left\{\frac{1}{T}\frac{L_{\rm Qs}}{\pi\Ds(T)}\right\}
       =\frac{1}{T}\left\{\frac{\bra \Je\Js
   \ket}{\Ds(T)}\frac{1}{T}-h\right\},
\label{Seebeck}
\end{equation}
are both finite at finite temperatures. 

\begin{figure}[hh]
\begin{center}
\includegraphics[width=0.45\textwidth]{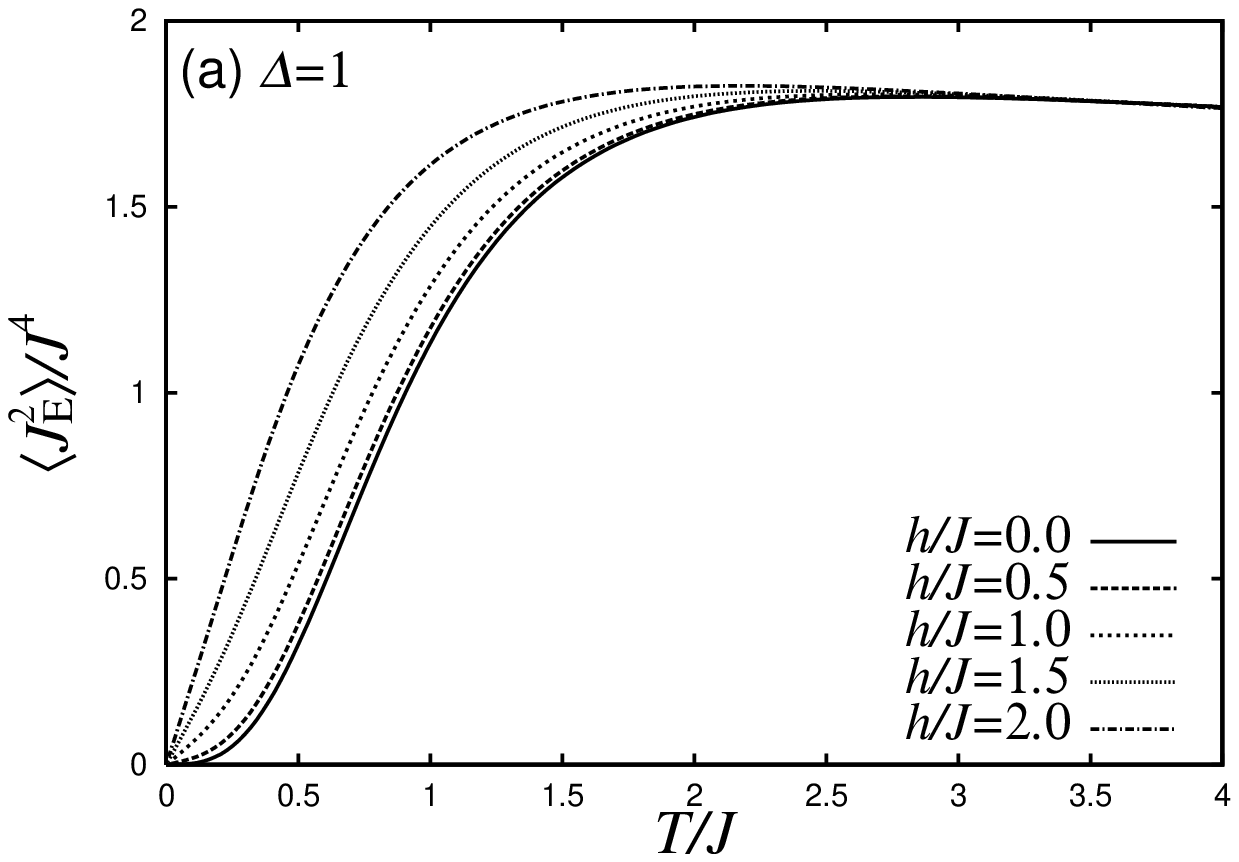}
\includegraphics[width=0.45\textwidth]{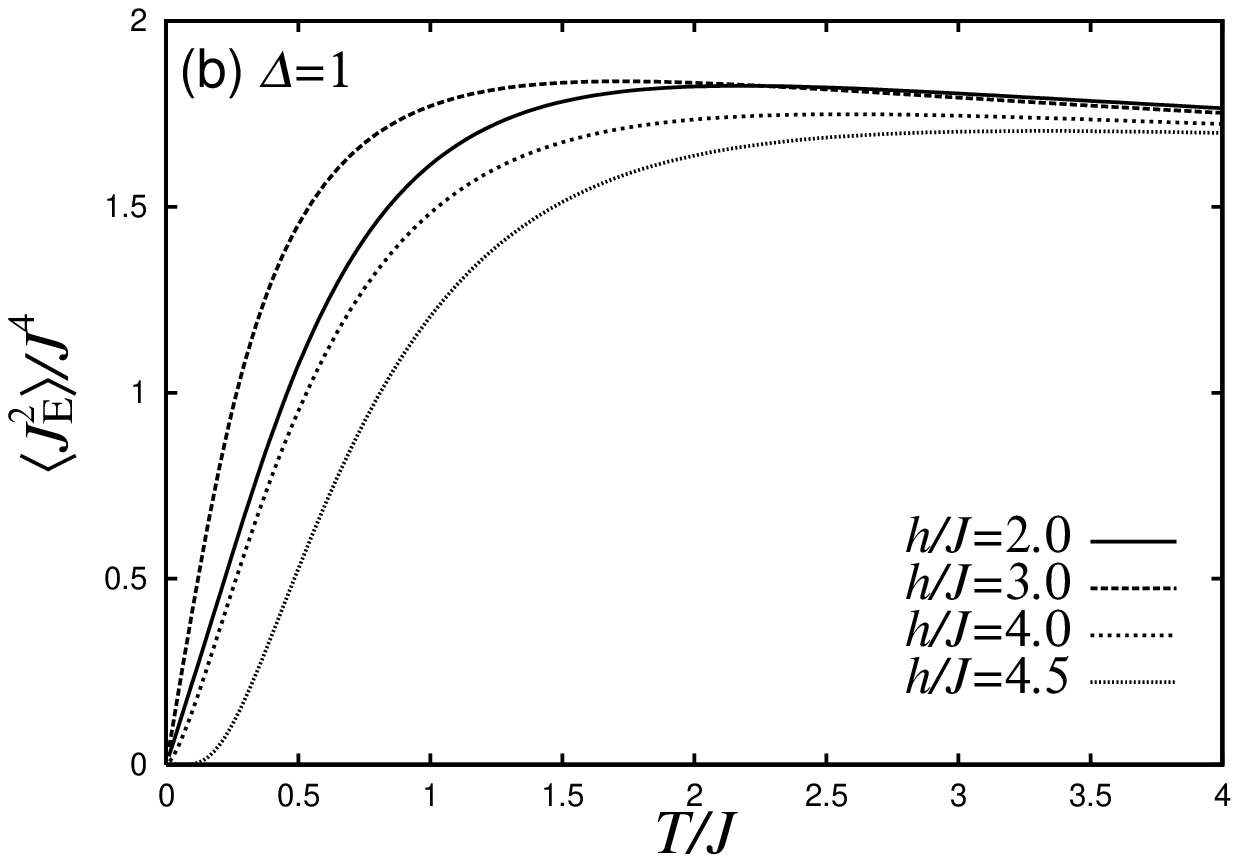}
\end{center}
\caption{Illustration of $\bra\Je^2\ket$ for the isotropic
case $\Delta=1$ as a function of temperature 
for various magnetic fields $h$.}
\label{f1}
\end{figure}

To obtain them, we have to evaluate the correlations
of $\Je$ and $\Js$. Let us consider the autocorrelation 
$\bra\Je^2\ket$ first.
To evaluate this quantity, we introduce the following 
extended Hamiltonian including $\Je$ as a perturbation;
$\tilde{\H}:=\lam_0 \H_0-h \M+\lam_1\Je$,  where $\H_0$ is
the Hamiltonian without the Zeeman term $h\M$. The parameters
$\lam_0$ and $\lam_1$ are introduced for later convenience
and should be taken to be $\lam_0=1$ and $\lam_1=0$ after
taking all necessary derivatives.
Introducing the partition function $Z=\tr \e^{-\beta\tilde{\H}}$,
we easily see that $\bra\Je^2\ket$ is 
evaluated by taking the second logarithmic derivative with respect 
to $\lam_1$: 
$\bra \Je^2 \ket=\partial_{\lam_1}^2\ln Z\big|_{\lam_1=0}/(L \beta^2)$.
Note that from symmetric arguments we can prove
$\bra \Je \ket=0$. Applying a lattice path integral formulation, 
we introduce the quantum transfer matrix (QTM) in the imaginary time 
direction. In this formalism $Z$ in the thermodynamic
limit $L\to\infty$ can be expressed as the largest eigenvalue of the 
QTM $\Lambda(\lam_0,\lam_1)$, namely $\lim_{L\to\infty}(\ln
Z)/L=\ln\Lambda(\lam_0,\lam_1)$:
\begin{equation}
  \ln \Lambda(\lam_0,\lam_1)=
          \frac{\beta h}{2}+\oint_{C}a_1(x+\i)\ln(1+\eta^{-1}(x+\i))\d x,
  \label{eigen}
\end{equation}
where the contour $C$ encloses the real axis (for instance we take 
$\Im x=\i$ ($\Im x=-\i$) for the upper (lower) contour) and 
$a_n(x)=\gamma\sin n\gamma/(2\pi(\cosh\gamma x-\cos n\gamma))$.
The unknown function $\eta(x)$ is determined by the following non-linear 
integral equation (NLIE):\cite{KlumTH2}
\begin{align}
\ln\eta(x)&=\beta(\lam_0+\lam_1 A \partial_x)e(x)+\beta h \nn \\
          &\quad +\oint_{C}a_2(x-y-\i)\ln(1+\eta^{-1}(y+\i))\d y,
  \label{nlie}
\end{align}
where $e(x)=-2 \pi A a_1(x)$ and $A=2J\sin\gamma/\gamma$.
Consequently we obtain
\begin{equation}
     \bra\Je^2\ket=\frac{1}{\beta^2}\partial^2_{\lam_1}
                   \ln\Lambda(1,\lam_1)\big|_{\lam_1=0}.
   \label{jth2}
\end{equation}
In Fig.~\ref{f1}, the temperature dependence of the
correlation $\bra\Je^2\ket$ for the isotropic point $\Delta=1$ is 
depicted for various magnetic fields. For $0<0<h_c$ ($h_c=4$ is
the critical magnetic field for $\Delta=1$), 
we find that $\bra \Je^2 \ket$ is linear in $T$ at low-temperature 
$T\ll 1$.
On the other hand, $\bra \Je^2 \ket$ 
exponentially decays for $h>h_c$ due to the mass gap.
At high temperature, $\bra \Je^2 \ket$ converges to a constant
$J^4(1+2\Delta)/2$ not depending on the magnetic field $h$.
%

Next we consider the correlation $\bra\Je\Js \ket$.
Since the spin current $\Js$ is not a constant of motion 
(except for the XY model $\Delta=0$), the quantity $\Je\Js$ is 
no longer conserved, namely $[\H,\Je\Js]\ne 0$. Hence the above 
procedure, which is useful to calculate the conserved quantities, 
is not applicable directly.
Alternatively we use the following non-trivial identity relating 
$\bra \Je\Js \ket$ to $\bra\Je^2 \ket$, which is valid for the
thermodynamic limit $L\to\infty$~\cite{LG03}
\[
\bra \Je \Js \Delta \mathcal{H}_0\ket=
\bra \Je^2 \Delta \mathcal{M}\ket, 
\]
where 
$\Delta \mathcal{H}_0=\mathcal{H}_0-\bra\mathcal{H}_0 \ket$
and $\Delta \mathcal{M}=\mathcal{M}-\bra\mathcal{M} \ket$.
Using this identity, we express $\bra\Je\Js\ket$ in
terms of the largest eigenvalue of the QTM $\Lambda(\lam_0,\lam_1)$:
\begin{equation}
   \bra\Je\Js \ket=-\frac{1}{\beta^2}\partial^2_{\lam_1}
              \int \partial_{h}
                      \ln\Lambda(\lambda_0,\lambda_1)\d\lambda_0
                           \biggl|_{\lambda_0=1,\lambda_1=0}.
    \label{jths}
\end{equation}
To simplify the above equation further, we take the derivative
with respect to $h$ and $\lam_0$ of both sides
of eq.~\eqref{nlie} after shifting the variable $x\to x+ \i$. 
Comparing the resultant equations, we find
\begin{align}
   &\oint_C (\partial_h\ln\eta(x+\i)-\beta)
             \partial_{\lam_0}\ln(1+\eta^{-1}(x+\i))\d x   \nn \\
    &=\oint_C (\partial_{\lam_0}\ln\eta(x+\i)-\beta e(x+\i))
               \partial_{h}\ln(1+\eta^{-1}(x+\i)) \d x. \nn
\end{align}
Therefore we quickly see
\[
\oint_C \partial_{\lam_0}\ln(1+\eta^{-1}(x+\i))\d x
       =-2\pi A \left\{
                  \partial_h \ln\Lambda-\frac{\beta}{2}
               \right\}.
\]
Here we have used $e(x)=-2\pi A a_1(x)$ and eq.~\eqref{eigen}.
Finally substituting the above equation into eq.~\eqref{jths},
we arrive at
\[
      \bra\Je\Js \ket=\frac{1}{2\pi A\beta^2}\oint_C
                \partial_{\lam_1}^2\ln(1+\eta^{-1}(x+\i))\d x.
\]

\begin{figure}[hh]
\begin{center}
\includegraphics[width=0.45\textwidth]{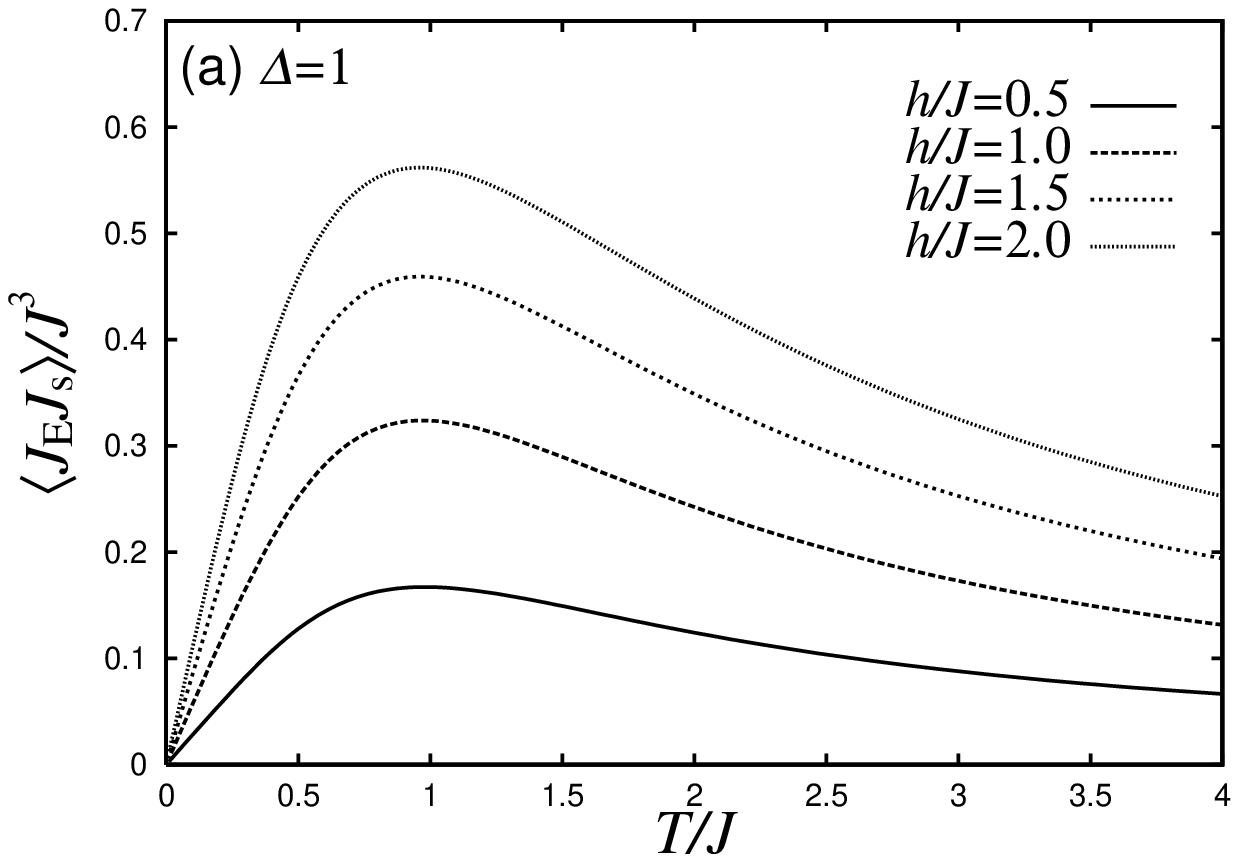}
\includegraphics[width=0.45\textwidth]{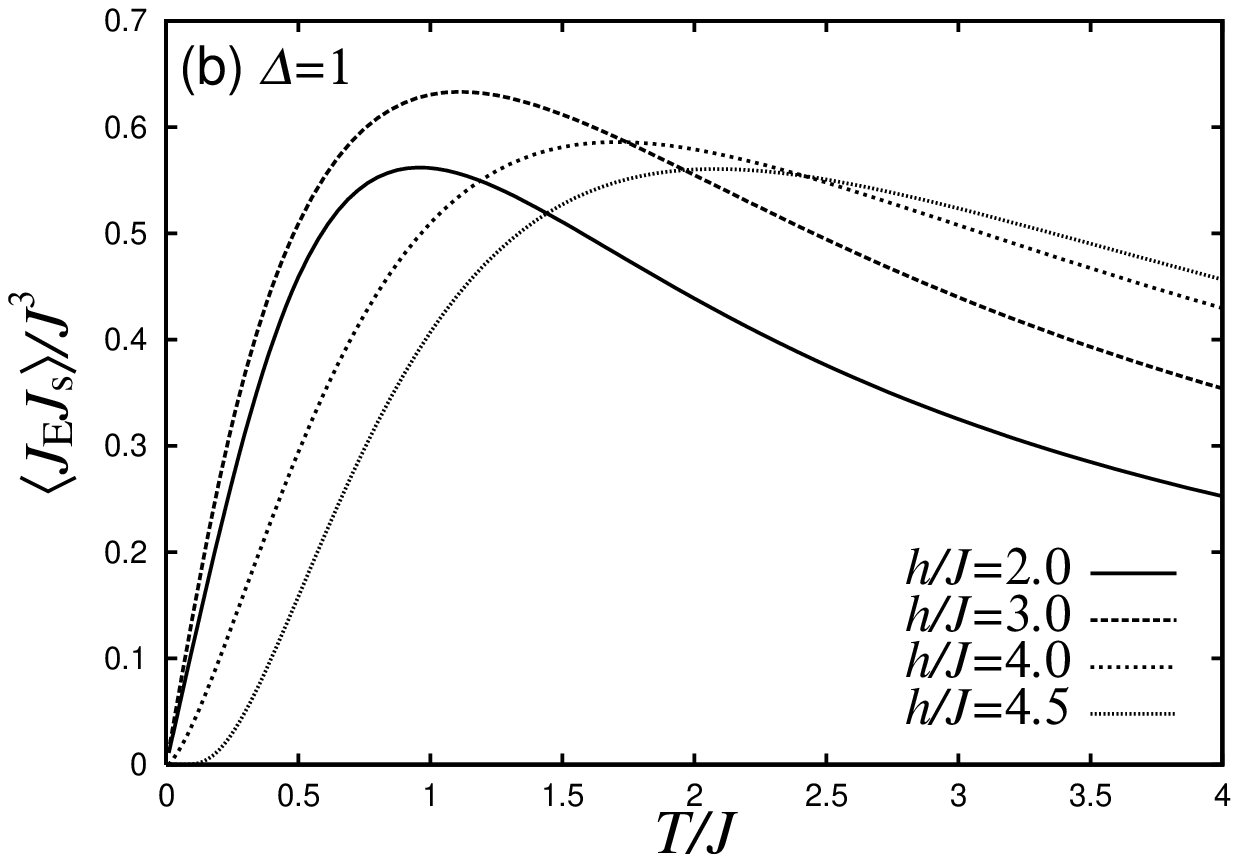}
\end{center}
\caption{Temperature dependence of the correlation
$\bra\Je \Js \ket$ of the isotropic case $\Delta=1$ for 
various magnetic fields.}
\label{f2}
\end{figure}

In Fig.~\ref{f2}, we show the temperature dependence 
of the current correlation $\bra\Je \Js \ket$ for the
isotropic point $\Delta=1$.
As mentioned above, this correlation is strictly zero 
at $h=0$.
At finite magnetic fields, we find that $\bra\Je\Js\ket$ 
has a finite temperature maximum which increases with 
increasing magnetic field for $0<h<h_c=4$. For
$h>h_c$, the maximum decreases and the corresponding 
temperature $T_0$ shifts to higher values with increasing
magnetic field. At $h\to\infty$ in which all spins 
point up, the temperature $T_0$ moves to
infinity for fixed $J$. Consequently $\bra\Je\Js\ket=0$
at $h=\infty$.
For $h$ weaker than the critical field $h_c$, $\bra\Je\Js\ket$ is linear in 
$T$ at low-temperature.
On the other hand, $\bra \Je\Js \ket$ exponentially 
decays for $h>h_c$ because of the existence of the mass gap.

\textit{Low-temperature asymptotics $(T\ll h)$ .--}
Next we would like to discuss the leading contributions
at low temperature $T\ll 1$ and $T\ll h$ (i.e. $\beta h\gg 1$) 
in which the ``logarithmic correction" terms especially for the 
isotropic case are next-leading contributions.
In this case the auxiliary function $\eta^{-1}(x)$
is exponentially small on the upper contour, i.e.
$\eta^{-1}(x+2\i)\sim \exp(-\beta h \delta)$ ($\delta>0$).
Moreover the function on the lower contour $\Im x=0$
behaves as $\ln(1+\eta^{-1}(x))\gg 1$ for the
region $x\in[\laml,\lamr]$ and $\ln(1+\eta^{-1}(x))\ll 1$
for $x<\laml$ and $x>\lamr$.
Therefore the NLIE \eqref{nlie} reduces to 
$\ln\eta(x)=\beta \varepsilon(x)+O(T)$, 
where $\ep(x)$ is the dressed energy given by
\begin{equation}
   \ep(x)=(\lam_0+\lam_1 A\partial_x)e(x)+h-
         \int_{\laml}^{\lamr}a_2(x-y)\ep(y)\d y.
   \label{denergy}
\end{equation}
The right (left) ``Fermi point" $\lamr$ ($\laml$)
depending on $\lam_0$ and $\lam_1$
is obtained from the condition $\ep(\lamr)=\ep(\laml)=0$.
To consider the $O(T)$ correction to $\ln\eta(x)$, we take into
account the behavior near the Fermi point: 
$\ln\eta(x)=\beta \ep^{\prime}(\Lambda_{\rm R/L})
(x-\Lambda_{\rm R/L})$ for $x\sim\Lambda_{\rm R/L}$.
Using this, we evaluate the $O(T)$ contribution
\begin{align}
&\(
  \int_{-\infty}^{\laml} \d y+\int_{\lamr}^{\infty} \d y
\)  a_2(x-y)\ln(1+\eta^{-1}(y)) \nn \\
&
\quad
+\(
\int_{\laml}^0 \d y+\int_{0}^{\lamr} \d y
\) a_2(x-y)\ln(1+\eta(y))
\label{orderT}
\end{align}
by changing the variable as $\ln \eta(y)\to z$
and using the values $\ln\eta(\lam_{\rm R/L})=0$
and $\ln \eta^{-1}(0)=\ln\eta(\pm \infty)=\infty$.
The result is
\begin{align}
\eqref{orderT}&\to
\frac{\pi^2}{6\beta}
  \left\{
     \frac{a_2(x-\Lambda_{\rm R})}{\varepsilon^{\prime}(\Lambda_{\rm R})}
    -\frac{a_2(x-\Lambda_{\rm L})}{\varepsilon^{\prime}(\Lambda_{\rm L})}
  \right\}. \nn
\end{align}
Thus we obtain the auxiliary function $\ln\eta(x)$ up to
$O(T)$:
\begin{align}
\ln\eta(x)&=\beta\varepsilon(x)
+\frac{\pi^2}{6\beta}
 \left\{
   \frac{a_2(x-\Lambda_{\rm R})}{\varepsilon^{\prime}(\Lambda_{\rm R})} -
   \frac{a_2(x-\Lambda_{\rm L})}{\varepsilon^{\prime}(\Lambda_{\rm L})} 
  \right\}.  \nn
\end{align}
Applying the same procedure and using the above obtained relations, 
we have the low-temperature behavior of the eigenvalue \eqref{eigen} 
up to $O(T)$:
\begin{align}
\ln\Lambda=
     \frac{\beta h}{2}&-
        \beta \int_{\laml}^{\lamr}a_1(x)\ep(x){\rm d}x \nn \\
&+\frac{\pi^2}{6\beta}
   \left\{
       \frac{\rho(\lamr)}{\varepsilon^{\prime}(\lamr)}
      -\frac{\rho(\laml)}{\varepsilon^{\prime}(\laml)}
    \right\}+O(T^2),
  \label{lT-eigen}
\end{align}
where $\rho(x)$ is the density function given by
\begin{equation}
    \rho(x)=a_1(x)-\int_{\laml}^{\lamr}a_2(x-y)\rho(y){\rm d}y.
  \label{density}
\end{equation}
Combining \eqref{lT-eigen} with \eqref{jth2} and \eqref{jths},
we obtain $\bra\Je^2\ket$ and $\bra\Je\Js\ket$ for $\beta h\gg 1$:
\begin{align}
&\bra \Je^2\ket=h^2 \Ds(0) T
+ \frac{(\pi A)^2T^3}{3}
    \biggl\{ 
        \frac{\alpha^{\prime}(\Lam)\rho(\Lam)}{\ep^{\prime 2}(\Lam)}
	+\frac{\alpha(\Lam)\rho^{\prime}(\Lam)}{\ep^{\prime 2}(\Lam)} \nn \\
   &\qquad\quad
     -\frac{2\alpha(\Lam)\ep^{\dprime}(\Lam)\rho(\Lam)}{\ep^{\prime 3}(\Lam)} 
     -\frac{\alpha(\Lam)}{2\pi A \ep^{\prime}(\Lam)}
     +\frac{\alpha^2(\Lam)\rho(\Lam)}{\ep^{\prime 3}(\Lam)}
    \biggr\}, \nn \\
&\bra \Je\Js \ket=h \Ds(0) T+\frac{\pi A T^3}{6}
 \biggl\{
       \frac{\alpha^2(\Lam)\xi(\Lam)}{\ep^{\prime 3}(\Lam)}
+       \frac{\alpha^{\prime}(\Lam)\xi(\Lam)}{\ep^{\prime 2}(\Lam)} \nn \\
&\qquad\quad
   +\frac{\alpha(\Lam)\xi^{\prime}(\Lam)}{\ep^{\prime 2}(\Lam)}-
    \frac{2\alpha(\Lam)\ep^{\prime\prime}(\Lam)\xi(\Lam)}{\ep^{\prime
   3}(\Lam)}
  \biggr\},
  \label{lTcorrelation}
\end{align}
where  $\Lam$ is the Fermi point determined by $\ep(\pm \Lam)=0$ 
(note that we set $(\lam_0,\lam_1)=(1,0)$ in \eqref{denergy} and 
\eqref{density});
$\Ds(0)$ is the zero temperature spin stiffness
$\Ds(0)=\vf \xi^2(\Lam)/\pi$; $\vf$ is the Fermi velocity 
defined by $\vf=\ep^{\prime}(\Lam)/(2\pi\rho(\Lam))$; the 
function $\alpha(x)$ and the dressed charge $\xi(x)$  are given by
\begin{align}
\alpha(x)&=e^{\prime\prime}(v)-\int_{-\Lam}^{\Lam}
                     a_2(x-y)\alpha(y)\d y, \nn \\
\xi(x)&=1-\int_{-\Lam}^{\Lam}a_2(x-y)\xi(y)\d y. \nn
\end{align}

\begin{figure}[hh]
\begin{center}
\includegraphics[width=0.45\textwidth]{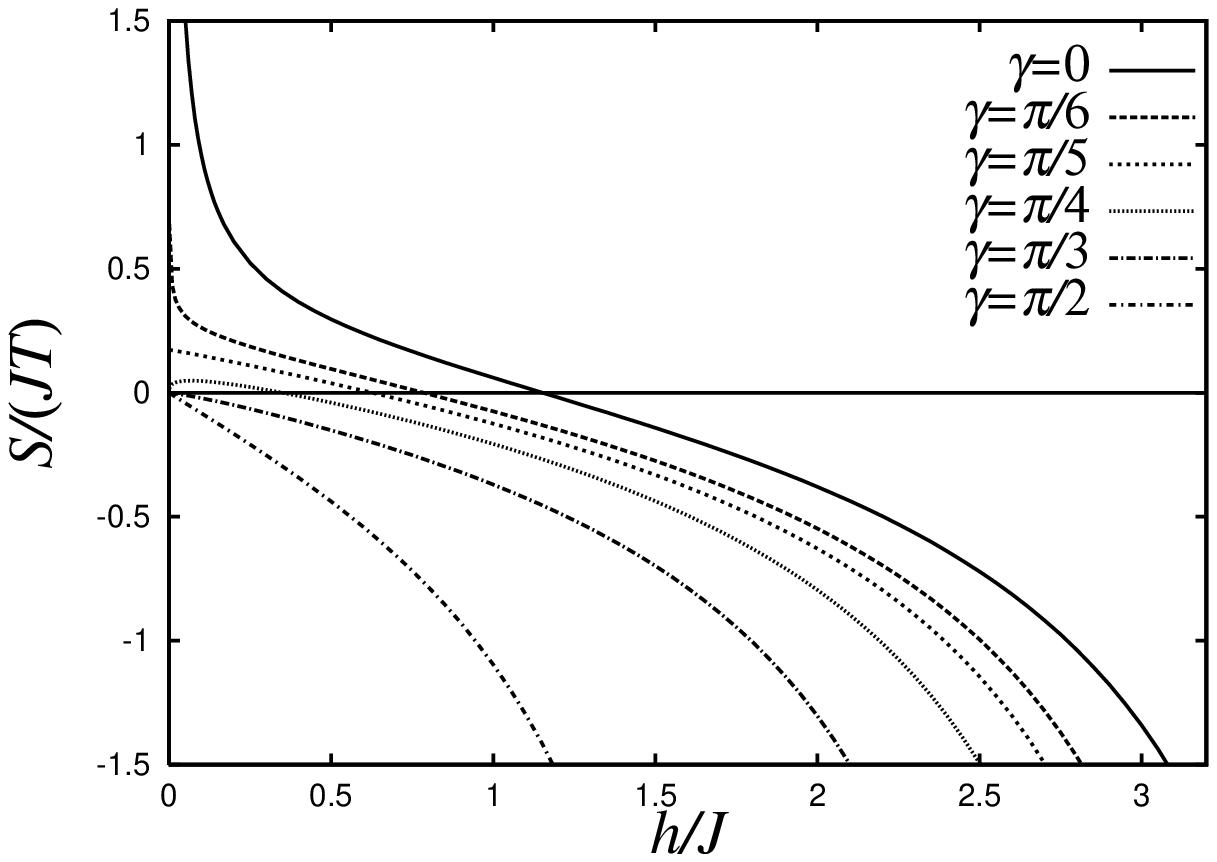}
\end{center}
\caption{
The leading contribution to the Seebeck coefficient at low temperatures
$T\ll h$.}
\label{f3}
\end{figure}

\textit{Drude weights and Seebeck Coefficient.}--
Finally we discuss the thermal Drude weight $\Dth(T)$
and the magnetic Seebeck coefficient
$S$ at low-temperature ($\beta h\gg 1$).
To evaluate them, we have to explicitly determine the spin Drude weight
$\Ds(T)$ as well as the above obtained current correlations $\bra\Je^2\ket$
and $\bra\Je\Js\ket$. 
Though the Drude weight for the spin transport at finite temperatures and zero
magnetic field has been evaluated by Zotos already a while ago\cite{Zotos99},
the validity of the results is still debated.
Here, we determine the low-temperature behavior alternatively without directly
calculating $\Ds(T)$.
Utilizing the phenomenological relation between
the thermal conductivity $\kappa$ and the specific
heat $C$, i.e. $\kappa=C(T)v_{\rm F}^2\tau$
($\tau$: relaxation time), we assume the low-temperature
asymptotics of the thermal Drude weight as
\begin{equation}
   \Dth(T)=\frac{\pi v_{\rm F}}{3}  T+O(T^2).
   \label{lTthermal}
\end{equation}
Here we have used $\kappa=\Dth(T)\tau$ and $C=\pi T/(3 \vf)$.
\cite{AlGros2} Note that $\tau=\infty$ for the present case. 
The validity of eq.\eqref{lTthermal} is verified at 
$h=0$~\cite{KS} and for the XY model ($\Delta=0$) for
$0\le h\le h_c$ ($h_c=2$).
Combining \eqref{lTthermal} with \eqref{thermalDrude} and
using the result \eqref{lTcorrelation}, we determine the 
low-temperature behavior of $\Ds(T)$. 
{}From the resultant equation and \eqref{lTcorrelation}, we obtain 
the ratio of $\bra\Je\Js \ket$ and $\Ds(T)$;
\begin{align}
\frac{\bra \Je\Js \ket}{\Ds(T)}=h T-\frac{\pi T^3}{6A\rho(\Lam)\xi(\Lam)}
      \left\{
          1+\frac{A \alpha(\Lam)}{2\pi \rho(\Lam)v_{\rm F}^2}
      \right\}. \nn
\end{align}
Finally substituting this into \eqref{Seebeck}, we arrive at
the leading low-temperature behavior of the magnetic Seebeck 
coefficient for $T\ll h$:
\begin{equation}
S(T)=-\frac{\pi T}{6A\rho(\Lam)\xi(\Lam)}
      \left\{
          1+\frac{A \alpha(\Lam)}{2\pi \rho(\Lam)v_{\rm F}^2}
      \right\} 
       +O(T^2). \nn
\end{equation}
In Fig.~\ref{f3}, the coefficient of the leading correction is 
depicted as a function of the magnetic field for various anisotropy 
parameters. For weak interaction strengths $\Delta<\Delta_0\sim 0.5$, 
the leading behavior is negative. On the contrary, for $\Delta>\Delta_0$, 
due to effects of strong correlations, the Seebeck coefficient changes 
sign at certain magnetic field $h_0$ (or equivalently certain
magnetization $\mathcal{M}_0$). The value $h_0$ shifts to higher values
with increasing the interaction strengths. Recently similar behavior 
has also been observed in the electric Seebeck coefficient of the 
Hubbard model for finite $U$ and $T$ by the numerical diagonalization for 
small systems.\cite{MP04}
%

In summary, we have discussed the magnetothermal
effects and the thermal transport at finite magnetic
fields. The magnetic Seebeck coefficient changes
sign above certain interaction strengths, 
because of effects of strong correlations.

\section*{Acknowledgment}
We would like to thank S. Fujimoto and
 X. Zotos for stimulating discussions.
This work was supported by a 21st Century COE Program at
Tokyo Tech ``Nanometer-Scale Quantum Physics" by the
Ministry of Education, Culture, Sports, Science and Technology.



\begin{thebibliography}{99}
\bibitem{ZP03} For a review see X.~Zotos and P.~Prelov\v sek:
  cond-mat/0304630.
%
\bibitem{Kudo} K.~Kudo, {\it et al.}: 
J. Low. Temp. Phys. {\bf 117} (1999) 1689.
%
\bibitem{Sologubenko00} A.~V.~Sologubenko, {\it et al}.: 
Phys. Rev. Lett. {\bf 84} (2000) 2714.
%
\bibitem{Hess01} C.~Hess,{\it et al}.: Phys. Rev. 
B {\bf 64} (2001) 184305.
%
\bibitem{Taki} M.~Takigawa,{\it et al}.:
Phys. Rev. Lett. {\bf 76} (1996) 4612.
%
\bibitem{znp} X.~Zotos, F.~Naef and P.~Prelov\v sek: Phys. Rev. 
B {\bf 55} (1997) 11029.
%
\bibitem{Zotos99} X.~Zotos: Phys. Rev. Lett. {\bf 82} (1999) 1764.
%
\bibitem{NarMA98} B.~N.~Narozhny, A.~J.~Millis and N.~Andrei: Phys. Rev. B
{\bf 58}  (1998) R2921.
%
\bibitem{AlGros1} J.~V.~Alvarez and C.~Gros: Phys. Rev. Lett. {\bf 88} (2002)
077203.
%
\bibitem{FK03} S.~Fujimoto and N.~Kawakami: Phys. Rev. Lett. {\bf 90} (2003)
 197202.
%
\bibitem{KS} A.~Kl\"umper and K.~Sakai:
J. Phys. A: Math. Gen. {\bf 35} (2002) 2173.
%
\bibitem{SK}  K.~Sakai and  A.~Kl\"umper:
J. Phys. A: Math. Gen. {\bf 36} (2003) 11617.
%
\bibitem{MHCB} F.~Heidrich-Meisner, et al.: 
 Phys. Rev. B {\bf 66}  (2002) 140406(R).
%
\bibitem{OCC03} E.~Orignac, R. Chitra and R.Citro: Phys. Rev. B
{\bf 67} (2003)  134426.
%
\bibitem{MHCB2} F. Heidrich-Meisner, {\it et al}.: 
Phys. Rev. {\bf B 68} (2003) 134436.
%
\bibitem{LG03} K.~Louis and C.~Gros: Phys. Rev. {\bf B 67} (2003) 224410.
%
\bibitem{MHCB3} F. Heidrich-Meisner, {\it et al}.: cond-mat/0408529.
%
\bibitem{Mahan} G.~D.~Mahan: {\it Many-Particle Physics} 
(New York: Plenum Press).
%
\bibitem{KlumTH2} A.~Kl\"umper: Z. Phys. B {\bf 91}(1993) 507.
%
\bibitem{AlGros2} J.~V.~Alvarez and C.~Gros: Phys. Rev. Lett. {\bf 89} (2002)
156603.
%
\bibitem{MP04} M.~M.~Zemljic and P.~Prelov\v sek: cond-mat/0407251.
%
\end{thebibliography}
\end{document}